\documentclass[sigconf, nonacm]{acmart}

\usepackage{balance} % for balancing columns on the final page
\usepackage{graphicx} % for pdf, bitmapped graphics files
\usepackage{xspace}
\usepackage{amsmath} % assumes amsmath package installed
\usepackage{bm}

\usepackage{enumitem}
\usepackage{hyperref}
\usepackage{nicefrac}
\usepackage{multirow}

\newcommand{\theGraph}{\textrm{$\mathcal{G}$}\xspace}
\newcommand{\nodeSet}{\textrm{$\mathcal{N}$}\xspace}
\newcommand{\edgeSet}{\textrm{$\mathcal{E}$}\xspace}
\newcommand{\agentSet}{\textrm{$\mathcal{A}$}\xspace}
\newcommand{\speed}{\textrm{\emph{$\bar{\Omega}$}}\xspace}

\newcommand{\theSchedule}{\textrm{$\mathcal{S}$}\xspace}

\newcommand{\maze}{\emph{maze-32-32-2}\xspace}
\newcommand{\Empty}{\emph{empty-16-16}\xspace}

\acmSubmissionID{443}

\title[]{The Cost of Executability: Temporal Margins for Continuous-Time Multi-Agent Path Finding under Closed-Loop Control}

\author{Sabino Roselli}
\affiliation{
  \institution{Chalmers University of Technology}
  \city{Gothenburg}
  \country{Sweden}}
\email{rsabino@chalmers.se}

\author{Alvin Combrink}
\affiliation{
  \institution{Chalmers University of Technology}
  \city{Gothenburg}
  \country{Sweden}}
\email{combrink@chalmers.se}

\author{Ze Zhang}
\affiliation{
  \institution{Agile Robots SE}
  \city{Munich}
  \country{Germany}}
\email{zhze.zhang@gmail.com}

\begin{abstract}
Multi-agent path finding (MAPF) solvers are evaluated on the schedules they produce, not on how well those schedules can be executed. 
Specifically, continuous-time optimal solvers can achieve even lower cost than their discrete counterparts by scheduling agents with near-zero temporal/spatial separation, and by exploiting graphs with higher connectedness; however, such schedules may prove hard to execute under a realistic controller. 
We integrate AOC-CBS, an anytime-optimal continuous-time MAPF solver, into a fleet coordination framework in which each agent tracks its timed path using model predictive control, and we evaluate the resulting system on congested MovingAI benchmark instances with 21 to 40 agents. Without intervention, only a fraction of the instances complete execution. 
We then apply $T$-robustness by adding a \emph{time-margin}, i.e., a minimum temporal gap between agents at shared locations, as well as spatial inflation of the agent footprint.
Increasing either margin raises execution success, for a relatively small increase in the schedule's sum-of-costs; it also improves how closely agents adhere to the schedule: when tracking each agent's position throughout execution and comparing it against the position it should occupy according to the schedule, the mean discrepancy falls significantly as the margin increases.

In general, the two margins have different effects; temporal margin decreases deviation, giving the controller room to stay on schedule. Spatial margin affects deviation less and instead increases the clearance between agents, so that the same deviation no longer leads to a blocking stall.
They are nonetheless substitutes rather than complements: spatial inflation is almost free in schedule cost, but its effect on execution success is small and does not compound with a temporal margin already in place.

Our results characterize a frontier between schedule optimality and execution reliability, and show where a fleet should operate on that frontier.

\end{abstract}


\newcommand{\BibTeX}{\rm B\kern-.05em{\sc i\kern-.025em b}\kern-.08em\TeX}

\begin{document}

%%% The following commands remove the headers in your paper. For final 
%%% papers, these will be inserted during the pagination process.

%\pagestyle{fancy}
%\fancyhead{}

%%% The next command prints the information defined in the preamble.

\maketitle 

%%%%%%%%%%%%%%%%%%%%%%%%%%%%%%%%%%%%%%%%%%%%%%%%%%%%%%%%%%%%%%%%%%%%%%%%
\newpage

\section{Introduction}

% As the use of mobile robots in several aspects of everyday life continues to spread, the need for \emph{better} scheduling algorithms increases; shorter computation times, extension to the realm of continuous time, inclusion of additional features such as multi-location tasks or heterogeneous agent models, are typical goals when researching the field of Multi-Agent Path Finding (MAPF). Regardless of the specific objective function chosen, an efficient solution typically results in a rather \emph{tight} schedule, i.e., a schedule in which the drivable areas are highly utilized. 

Coordinating a fleet of mobile robots in a shared workspace is commonly decomposed into two layers~\cite{parker2016multiple}. 
A high-level scheduler solves a Multi-Agent Path Finding (MAPF) problem, assigning each agent a conflict-free schedule defining how it should move within a graph representation of the space. 
A low-level controller then drives each agent to follow its schedule as closely as the agents' movement dynamics allow.
This decomposition is attractive because it separates a discrete combinatorial problem from a continuous one: the scheduler reasons about who goes where and in what order, while the controller reasons about specific agent motion models and shapes, acceleration limits, and obstacle clearance.

The two layers are, however, often developed and evaluated in isolation. 
When looking at the high-level scheduling, the MAPF literature commonly measures a solver by its success rate of finding schedules within a time budget, and how well the schedules move agents to complete their tasks~\cite{stern2019multi}.
Motion models and shapes are often approximated in MAPF to various degrees.
The common classical, or discrete-time, MAPF formulation assumes that each edge takes exactly one time unit to traverse and agents do not conflict if separated on the graph~\cite{stern2019multi}.
Modern solvers~\cite{CBS, li2019disjoint, li2019symmetry, LaCAM} for this problem can plan for hundreds to thousands of agents.
Various works have aimed to extend discrete-time MAPF to consider motion models both during scheduling~\cite{ma2019lifelong, yan2024multi, yan2025multi} and after as a post-processing step~\cite{honig2016multi}.
\emph{Continuous-time} MAPF~\cite{andreychuk2022multi} 
encompasses several formulations allowing for real-valued time, with several recent algorithms~\cite{cohen2019optimal, andreychuk2022multi, OCCBS, combrink2026aoc} considering motion models and agent shapes during scheduling.
However, all of these works still assume that agents are perfectly able to follow the schedules.

To our knowledge, only a few studies~\cite{moldagalieva2026db, ConcreteMAPP} evaluate whether the resulting schedule can actually be executed.
% Which objective to minimize shapes the schedules produced, and for MAPF the this remains an active question~\cite{gabay2026reachability}. 
% However, under any of the common objectives, optimality is achieved by removing slack.
% A solver minimizing delays has an incentive to schedule two agents through a shared region without temporal and spatial separation.
%-----
Many of the common objective functions aim to minimize the time for an agent to complete its task.
Under this premise, slack is always removed if that entails a lower objective value.
Thus, MAPF algorithms under these objective functions are incentivized to schedule two agents through a shared region without temporal and spatial separation.  
%----- AC
This is harmless if the agents follow their schedules perfectly; 
however, under real-world conditions, that is seldom possible. 
A real robot decelerates into a turn, accelerates out of it, and tracks its reference with a non-zero error that depends on its dynamics and on the controller's efficacy. 
A schedule whose efficiency depends on two agents passing without margin from one another has no capacity to absorb that error. 
The first small delay propagates, and what was a sound schedule fails in execution.
%-----
% A schedule without margin between agents that is executed with a non-zero error risks leading to collisions.
%----- AC

This issue is acknowledged in the literature~\cite{yan2025analyzing, dunkel2024streamlining}, and it has been addressed in different ways. 
On the one hand, robust MAPF formulations, such as k-robust planning~\cite{atzmon2020robust} and its continuous counterpart T-robust~\cite{tan2024robust}, build tolerance into the schedule by ensuring the schedule remains valid under any subset of agents being delayed by at most k or T time units. 
On the other hand, execution-management approaches, notably the Action Dependency Graph~\cite{honig2019persistent}, preserve the ordering implied by the schedule and let agents wait for one another; later work switches that ordering online as delays accumulate~\cite{berndt2020feedback} and streamlines its construction~\cite{dunkel2024streamlining}.

Both directions acknowledge that robust schedules are more conservative than nominal ones. Neither quantifies how robust a schedule should be in a continuous-time setting with a controller actually closing the loop.

\begin{figure*}[ht]
    \centering
    \includegraphics[width=\linewidth]{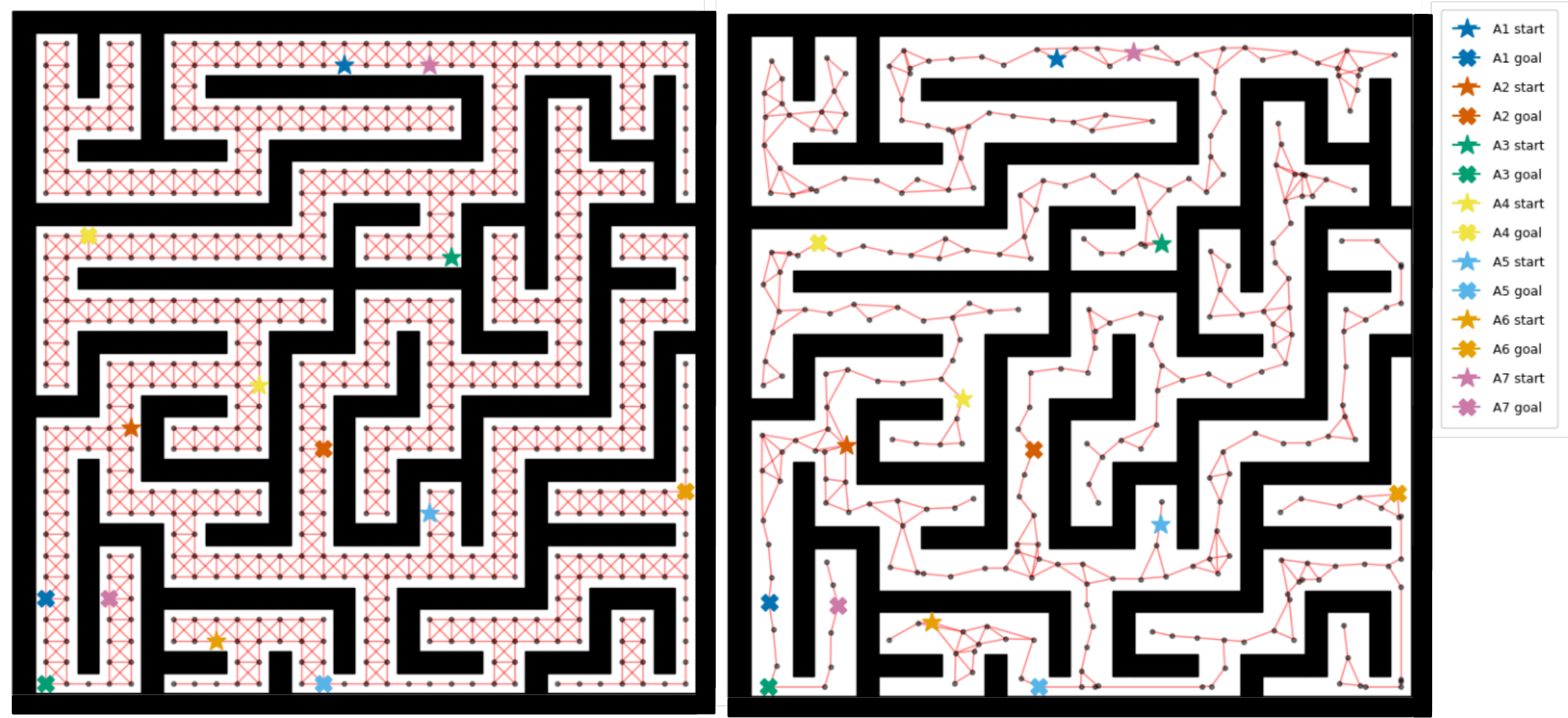}
    \caption{Visualization of the \maze map with 7 agents and their start and goal nodes. On the left, the grid-like graph with 8-connectedness is overlapped with the map; on the right, the alternative sampling method with node density equal to $0.1$.}
    \Description{Visualization showing a maze map with 7 agents, their start and goal nodes, and two sampling methods.}
    \label{fig:PaperExample}
\end{figure*}

In this work, we integrate AOC-CBS~\cite{combrink2026aoc}, an anytime-optimal algorithm for continuous-time MAPF, into a fleet coordination framework in which each agent tracks its schedule using model predictive control, 
and we execute the resulting schedules in simulation on congested instances drawn from the MovingAI benchmark set~\cite{stern2019multi}.
We build on our previous work~\cite{roselli2026combining}, in which we first presented the fleet coordination framework, where we adopted ComSat~\cite{roselli2024conflict} as a scheduling solver and tested its feasibility on two synthetic problem instances.
Replacing ComSat with AOC-CBS provides greater scalability but, most importantly, it can produce schedules with motions closer to the agent's motion models.

% \vspace{3mm}

% \textbf{[missing paragraph to preserve anonymity]}

% \vspace{3mm}

% In fact, discrete-time algorithms assume that all actions have uniform duration and that time advances in synchronous timesteps~\cite{stern2019multi}. 
% This forces every edge to have the same traversal time, which restricts these methods in practice to \emph{4-connected grids}: richer neighborhoods introduce moves of irrational duration relative to the cardinal one (a diagonal step
% costs $\sqrt{2}$) which no uniform timestep can represent, so agents cease to move in lock-step~\cite{walker2018extended, wu2026cbsaa}. 
% ComSat solves the discrete-time MAPF problem with unit-length edges,
Discrete-time algorithm solve the MAPF problem with unit-length edges,
therefore in practice restricting the graph to a grid-like structure, such as the common \emph{4-connected} grid with only cardinal movements.
Additionally introducing diagonal movements (\emph{8-connectedness}) with length $\sqrt{2}$ causes agents to no longer move in lock-step~\cite{walker2018extended, wu2026cbsaa}.
Continuous-time formulations remove this restriction, admitting actions of arbitrary duration, agents with geometric extent, and general graphs including sampled roadmaps~\cite{andreychuk2022multi}.

We evaluate the upgraded framework over two maps of the MovingAI dataset, increasing occupancy density until agents begin stalling due to congestion. 
We then investigate the effects of artificially increasing the slack in the system in two different ways; first, by increasing the size of the agents, and then by introducing a time margin, i.e., a minimum temporal gap between agents at shared locations. Sweeping these parameters traces a frontier between schedule cost and execution reliability.

Our findings are threefold. First, with no margin, only a fraction of instances are successfully executed. Some runs are interrupted by collisions, but most instead exceed the lateness threshold; at this operating point, mean schedule deviation, i.e., the distance between an agent's logged position during execution and the position it should occupy according to the schedule, reaches $3.13$~m on the maze. 
Second, a temporal margin is beneficial. The smallest margin we test ($m=2$) raises execution success to 41\% and 78\% for \maze and \Empty, respectively, for a 4 to 6\% increase in the schedule sum-of-costs.
Larger margins continue to pay in constrained environments, but returns diminish, and the limit becomes the scheduler rather than the controller; on the 4-connected maze graph AOC-CBS solves 92\% of instances at $m=10$ margin and only 23\% at $m=30$, whereas the 8-connected graph, whose richer neighborhood leaves the solver more routing freedom, remains fully solvable throughout. 
Third, spatial inflation costs almost nothing in schedule quality. On the other hand, it leaves schedule deviation essentially unchanged its effect on execution success is significant in only 6 of 72 paired comparisons between radii at a fixed margin, helping on the \Empty when no temporal margin is present, and negatively affecting the result on the \maze when combined with a margin.

The contribution is a characterization, on standard benchmarks and with a controller in the loop, of what schedule-level optimality costs at execution time, and of 
%the AC - removed since there are other mechanisms too. 
two mechanisms available for buying it back.

%%%%%%%%%%%%%%%%%%%%%%%%%%%%%%%%%%%%%%%%%%%%%%%%%%%%%%%%%%%%%%%%%%%%%%%%

\section{Preliminaries} \label{sec:problem_definition}

In this section, the fleet management framework is divided into a scheduler and a controller. We formalize the problem and provide the interface for each module. 

\textbf{The Environment and the Input}:
The workspace is assumed to be 2-dimensional.
Figure~\ref{fig:PaperExample} shows one of the two 2D maps (\emph{Maze-32-32-2}) from MovingAI~\cite{stern2019multi} used in this work; dark areas are not drivable or, in other words, are static obstacles, while blank areas are drivable.

Maps are abstracted into a finite, strongly connected, weighted, directed graph $\theGraph(\nodeSet,\edgeSet)$, with node set $\nodeSet$ and edge set $\edgeSet$. 
The abstraction is performed using two different techniques that result in graphs with different structures, 
one grid-like (Figure~\ref{fig:PaperExample} left), and one roadmap sampled from the drivable area (Figure~\ref{fig:PaperExample} right); 
more details about instance generation are found in Section~\ref{sec:experiments}.

The problem is also characterized by a set of start and goal locations, corresponding to specific nodes in the graph, one for each agent in the instance. For example, in an instance with $x \in \mathbb{N}$ agents, the agent set would be $\agentSet = \{(s_1,g_1),\ldots,(s_x,g_x)\,|\,s_i,g_i \in \nodeSet, \forall i\}$.
% where $s_i,g_i \in \nodeSet$ for all $i=1,\dots,m$.

\textbf{Scheduler}:
% The schedule is computed over a finite time interval $[0,\timeHorizon]$, where the \emph{Schedule Horizon} \timeHorizon is a fixed and real-valued instant when all agents should arrive at their goal. 
During scheduling, it is assumed that agents travel at constant speed \speed. Traversing a straight edge $e\!=\! (n,n')\!\in\!\edgeSet$ takes duration $|e| / \speed$, since traversals are not interrupted until the agent reaches the next node. 
An agent may begin a traversal at any real-valued time and may wait at a node for any real-valued duration. 
%An agent that has reached its goal remains there.  AC: sharpened a bit.
An agent completes its task by reaching its goal and remaining there indefinitely.  
Its individual schedule is therefore required to be conflict-free not only until it arrives but for all later time, which rules out solutions in which an agent completes its own route but then blocks another.
This matters in narrow segments, where a stationary agent cannot be passed.

Unlike the discrete-time MAPF, in which every action (wait or traversal) has the same duration and time advances in synchronous steps, in continuous time, the duration of a \emph{move} action follows from the length of the road segment, so graphs with edges of different lengths are admissible~\cite{andreychuk2022multi}.

The scheduler accepts the graph \theGraph and the agent set \agentSet, 
and produces a \emph{schedule} \theSchedule containing one \emph{individual schedule} $s_a$ for each agent $a \in \agentSet$.
The cost of an agent's individual schedule is the time at which the agent reaches its goal and remains there.
The cost of the schedule is the so-called sum-of-costs (SOC), being the sum over all individual schedules.
A good schedule reduces the SOC, and an optimal schedule minimizes it.

\textbf{Controller}:
For each agent, we generate a \emph{local reference trajectory} at every step. For agent $a$, we query the schedule \theSchedule for the individual schedule $s_a$ and generate a local reference trajectory $\mathcal{T}^{(a)}_k$ at each time step $k$, consisting of a sequence of $N$ points, where $N$ is the predictive horizon for the MPC controller.

For each agent $a$, the controller takes the reference trajectory $\mathcal{T}^{(a)}_k$ and computes a collision-free action $\bm{u}^{(a)}_k$ accordingly at time step $k$. The controller should avoid collisions with both environmental obstacles and other agents.

\subsection{Scheduler} \label{sec:fleet_scheduler}

% We solve this problem with AOC-CBS~\cite{combrink2026aoc}, which addresses a more general setting than the one above. Agents may follow sequences of goals, occupy vertices for prescribed service durations, and move on individual graphs; however, we restrict the instances to have one single goal per agent. 
Scheduling is done here using AOC-CBS~\cite{combrink2026aoc}, a continuous-time MAPF solver that early on produces an incumbent solution and progressively refines it over runtime, to eventually converge on an optimal solution.

AOC-CBS is based on the exact continuous-time MAPF solver OC-CBS~\cite{OCCBS}, which in turn is based on Conflict-Based Search (CBS)~\cite{CBS}.
Like CBS, AOC-CBS searches on two levels: 
a low-level that finds the lowest-cost path for a single agent satisfying a set of constraints, and a high-level over a binary constraint tree (CT).
Every CT node $N$ holds a set $N_{\mathcal{C}}$ of conflict-avoidance constraints and a candidate solution $N_{\Pi}$ containing one path for each agent, each path satisfying $N_{\mathcal{C}}$.
At every iteration, the CT node with the lowest-cost candidate solution is selected; 
if it contains no conflicts (i.e., it is a solution), then it is guaranteed to be the optimal solution and is returned;
otherwise, a conflict is selected and used to create two child CT nodes, each containing an additional constraint for one of the agents involved in the conflict.
The CT starts with only the root $R$ where $R_{\mathcal{C}}\!=\!\varnothing$ and progressively expands the tree, constraints being accumulated with depth, until a solution is found.
OC-CBS extends CBS to continuous time, and AOC-CBS extends OC-CBS by expanding several nodes at a time, with each expansion applying repair functions to find incumbent solutions.
Thus, AOC-CBS searches for the optimal solution while at the same time providing an incumbent solution with a known upper bound on the optimality gap. 

\subsubsection{Conflict Definition and Robustness Margins}\label{sec:margins}
 
AOC-CBS takes the conflict relation as an external predicate: given two agents, each in some state, the predicate decides whether the pair conflicts. We use this to introduce robustness into the schedule without modifying the search.
 
Let $z_a(t) \in \mathbb{R}^2$ be the position of agent $a$ at time $t$ under a given schedule. We parameterize the predicate by a radius $r$ and a temporal margin $m$, and declare agents $a$ and $a'$ to conflict whenever 
\begin{equation}
  \exists\, t,t' \;:\;
  |t - t'| \le m
  \;\wedge\;
  \lVert z_a(t) - z_a'(t') \rVert < 2r .
  \label{eq:conflict}
\end{equation}
%----- ORIGINAL VERSION
% With $m = 0$, this is the nominal predicate, under which agents conflict exactly when their footprints overlap at the same instant. Increasing $r$ inflates the footprint the scheduler plans with, producing schedules that keep agents further apart in space. Increasing $m$ requires agents to be separated in time.
 
% The temporal margin is the continuous-time counterpart of $k$-robustness, under which a schedule must remain valid given up to $k$ time-unit delays per agent.
% Our purpose is not to compute such schedules more efficiently but to measure how robust they are once a controller executes them, and what they cost. 
 
% Both parameters affect the outcome similarly, one in space and one in time, which makes them directly comparable as instruments for the same purpose. Section~\ref{sec:experiments} sweeps them separately and together.
%----- AC's VERSION
The agent's spatial footprint is defined by $r$; increasing $r$ inflates this footprint and enforces a larger spatial separation.
Increasing $m$ enforces a temporal separation of the agents' footprints, with $m=0$ being AOC-CBS's nominal predicate.
The temporal margin defined by $m$ is equivalent to T-robustness~\cite{tan2024robust}. 
Both parameters, $r$ in space and $m$ in time, provide instruments for the same purpose of incorporating robustness under non-zero execution error.
Section~\ref{sec:experiments} sweeps them separately and together.
%-----

\subsection{Bridging the Scheduler and Controller} \label{sec:the_bridge}

The scheduler assumes that agents travel at constant speed, accelerate and decelerate instantly, and take no time to turn. 
In reality, agents may accumulate delays due to several factors, including, but not limited to, the above-mentioned 
assumptions not holding for the real robots. 
This makes it practically infeasible to directly follow the schedule with basic schedule-tracking controllers. 
To address this, we need to bridge the global scheduler with the local controller, dynamically adjusting the reference speed and reference trajectory of each agent according to the given schedule.

\begin{figure}[t]
    \centering
    \includegraphics[width=\columnwidth]{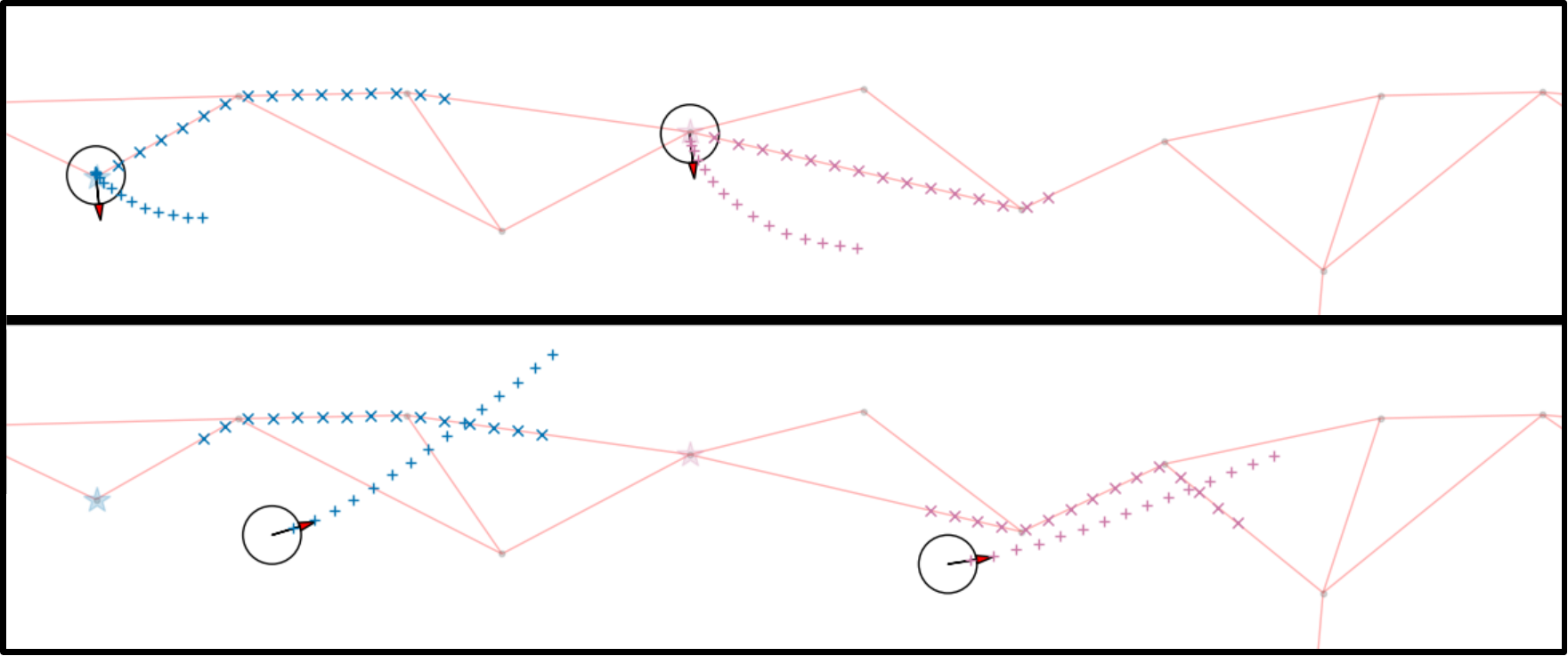}
    \caption{Detail of the map \maze picturing two agents at time $0$ (above), and $3$ seconds later (below). For each agent, $\times$ marks the reference trajectory, while $+$ marks the MPC output.}
    \Description{Detail of the map showing two agents at two different times, with $\times$ marking the reference trajectory and $+$ marking the MPC output.}
    \label{fig:Zoom}
\end{figure}

During execution, for each agent $a$, its individual schedule is a global reference path 
%from the start point ($x,y$ coordinates in the 2D map, corresponding to the start node $s_a$), 
in 2D coordinates from $a$'s start, % AC
passing through all intermediate nodes, to the end point. 
To achieve real-time performance, a global reference trajectory is precomputed upon receiving the schedule by sampling the reference path under the assumption of maximal velocity $\Omega_{\max}$, rather than generating a local trajectory from scratch at each time step. During operation, an initial reference trajectory $\bar{\mathcal{T}}_k$ is extracted from the global trajectory based on the current location of the agent. Assuming the current time is $t$ and the scheduled arrival time for the next node is $t_n$, the desired speed for the agent at time step $k=t/\Delta t$ will be
$$
\Omega_k = \min\left(\Omega_{\max},\, L_{\text{next}} / \max(0, (t_n-t))\right),
$$
where $L_{\text{next}}$ is the distance to the next node. Then, a new reference trajectory $\mathcal{T}_k$ (\emph{x} in Figure~\ref{fig:Zoom}) is generated by resampling the initial one $\bar{\mathcal{T}}_k$ with the ratio $\Omega_k/\Omega_{\max}$.

\subsection{Model Predictive Controller}\label{sec:MPC}

For agent $a$, after obtaining the desired speed $\Omega^{(i)}_k$ and reference trajectory $\mathcal{T}^{(i)}_k$, the MPC problem can be solved (generating the $+$ marks in Figure~\ref{fig:Zoom}.
The MPC objective has three parts: reference tracking, obstacle avoidance, and fleet collision avoidance. For reference tracking, at time step $k$,
\begin{equation}
    J_R(k) = || \bm{s}_{k}-\tilde{\bm{s}}_{k}||^2_{\bm{Q}_s} 
    \!+ ||\bm{u}_{k}\!-\!\tilde{\bm{u}}_{k}||^2_{\bm{Q}_u} 
    \!+ ||\bm{u}_{k}\!-\!\bm{u}_{k-1}||^2_{\bm{Q}_a},
\end{equation}
where $\bm{s}_{k}$ is the state, $\tilde{\bm{s}}_{k}$ is the reference state from $\mathcal{T}^{(i)}_k$, $\bm{u}_{k}$ is the action, $\tilde{\bm{u}}_{k}$ is the reference action composed by $\Omega^{(i)}_k$, and all $\bm{Q}$ variables are corresponding weight matrices. The obstacle avoidance term $J_{\mathcal{O}}(k)$ is based on the indication of the agent being inside or outside of a polygonal obstacle \cite{cendismpc_2022}. The fleet collision avoidance is defined as
\begin{equation}
    J_F(j) = \max\left[0, Q_f\cdot\left(d_{\text{fleet}}-||\bm{z}_{k}^{(i)}-\bm{z}_{k}^{(j)}|| \right)^2\right],
\end{equation}
where $\bm{z}^{(a)}$ and $\bm{z}^{(a')}$ are the positions of agent $a$ and $a'$ respectively, $d_{\text{fleet}}$ is the safe distance between agents, and $Q_f$ is the weight parameter.

Assuming the current time step $k=0$, the complete MPC problem can be formulated as
\begin{align}
    \min_{\bm{u}_{0:N-1}} & \sum_{{k}=0}^{N-1} \bigg[ J_R(k) + J_{\mathcal{O}}(k) + \sum_{j=1,j\ne i}^{|\agentSet|}J_F(j) \bigg],  \\
    \text{s.t.} \quad 
    & \bm{s}_{k+1} = f(\bm{s}_{k}, \bm{u}_{k}), \, \forall k = 0, \ldots, N-1, \\
    & \bm{u}_{k} \in [\bm{u}_{\text{min}}, \bm{u}_{\text{max}}], \, \forall k = 0, \ldots, N-1, \\
    & \Dot{\bm{u}}_{k} \in [\Dot{\bm{u}}_{\text{min}}, \Dot{\bm{u}}_{\text{max}}], \, \forall k = 0, \ldots, N-1, \\
    & \bm{z}_{k} \notin \mathcal{O}, \, \forall k = 0, \ldots, N-1,
\label{eq:mpc}
\end{align}
where $f(\cdot)$ is the motion model, $\Dot{\bm{u}}_{k}$ is the derivative of the action, and $(\bm{u}_{\text{min}/\text{max}}, \Dot{\bm{u}}_{\text{min}/\text{max}})$ are physical limits of the agent's action.

In this work, each agent's motion model is a differential-drive with state $\bm{s}=(x,y,\theta)$ and action $\bm{u}=(v,\omega)$, following the unicycle model $\dot x = v\cos\theta$, $\dot y = v\sin\theta$, $\dot\theta = \omega$, discretized with fourth-order Runge--Kutta at $t_s=0.2$\,s. The action is bounded by $v\in[-0.1,1.5]$, $|\omega|\leq0.5$, $|\dot v|\leq1$ and $|\dot\omega|\leq3$.
%%%%%%%%%%%%%%%%%%%%%%%%%%%%%%%%%%%%%%%%%%%%%%%%%%%%%%%%%%%%%%%%%%%%%%%%

\section{Experimental Setup}\label{sec:experiments}

This section presents the experimental setup.
We used the two maps \maze and \Empty from the MovingAI dataset~\cite{stern2019multi},
along with their respective agent start-goal pairs from \emph{random scenario 1}.
%For both maps, we use \emph{random scenario 1} as input for the agents' start and goal locations.
These two maps were picked as they represent two opposite ends of a range, with \maze being a constrained environment with narrow corridors, and \Empty being an empty space. 
The idea is for the MPC to resolve local conflicts and compensate for agent delays;
%this ability is hindered by the presence of static obstacles~\cite{qian2026qdmapper}, 
static obstacles near local conflicts complicate their resolution,
and by choosing these maps, we can evaluate executability in two extreme contexts. 

The MovingAI maps are divided into cells, with each cell being either drivable (white) or not (black). 
These cells are used to generate the graph that the scheduler takes as input. 
The common way to do so is to place a node at the center of each drivable cell and connect them based on their proximity to each other (see Figure~\ref{fig:PaperExample} left). Nodes can be connected in a grid-like fashion with only cardinal connections (4-connectedness), but also additionally with diagonals (8-connectedness). 
With 4-connectedness, all edges in the graph have the same length, and all turns the agents perform are 90 or 180 degrees. 
On the other hand, 8-connectedness relaxes this constraint by introducing the diagonals, longer than cardinal edges, and giving rise to additional possibilities to turn at 45 degrees. 
% Ideally, a smaller turning radius might be easier and faster to handle by the MPC. 
With agent motion models with limited turning radius, the milder turn angles that diagonals offer should be easier to follow.

In addition to grid-like graphs, we also sample \emph{roadmaps} using the method described in~\cite{combrink2026aoc}, which takes as input a desired node density and samples nodes within the map's free space (see Figure~\ref{fig:PaperExample} right). 
Using this method allows us to generate graphs with more variation in the edge lengths and in the turning angles.

\begin{figure*}[h]
  \centering
  \includegraphics[width=\textwidth]{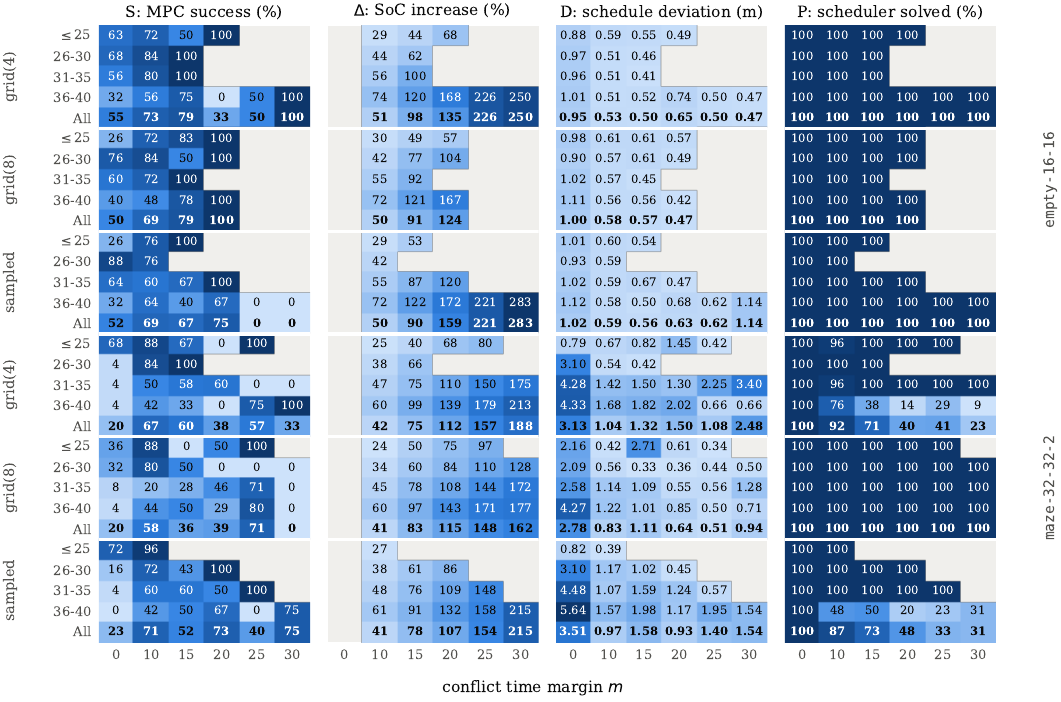}
  \caption{Effect of the conflict time margin $m$ ($r=0.35$). S = MPC success over instances the scheduler solved; $\Delta$ = increase of the planned sum of costs w.r.t.\ the $m=0$ baseline on the same instance; D = mean schedule deviation; P = share of instances the scheduler solved within the 60\,s budget. Darker is larger, each panel on its own scale; grey = not run. Margins $\geq15$ were run only where $m=0$ failed, so S and D are not averaged over the same instances across columns; $\Delta$ is zero at $m=0$ by construction. Bold rows aggregate 21--40 agents.}
  \Description{}
  \label{fig:heat_margin}
\end{figure*}

For our experiments, we also set the size of each cell to $2$, basically scaling up the entire map by a factor of 2. We empirically tried different values of ratios between agent and map size and eventually landed on a baseline agent radius $r= \nicefrac{\sqrt{2}}{4}$ and a cell size equal to $2$. This is because larger ratios, i.e., larger agents compared to the map size, led to agents getting stuck too easily as the MPC tries to maintain a safety distance from static and dynamic objects (the other agents).

Early stage experiments showed that, for both maps, there was almost never any congestion problem with fewer than $20$ agents running simultaneously. For this reason, we ran tests with a number of agents ranging from $21$ to $40$.
As mentioned before, we sweep over different values of agent radius and time margin, with specific values depending on the experiment batch. For each combination of the above-mentioned parameters, we generate five instances with a different random seed that affects the generation of the graph.

Finally, the scheduler assumes speed $\speed = 1$, meaning that agents span one unit length in one unit of time; during execution, however, the agents can reach a max speed of $\speed_{max} = 1.5$. This allows the agents to compensate for delays due to non-negligible turning time or management of encounters with other agents.

All experiments were run on an AMD Ryzen 9 9950X3D 16-Core Processor with 126GB of RAM, running Ubuntu 26.04 LTS. The implementation of the framework, including AOC-CBS and the MPC, as well as the code to generate the instances and run the experiments, is available at
% \textbf{[repository link omitted for anonymity]}.
\url{https://github.com/Chalmers-Control-Automation-Mechatronics/TrajPlan-ScheMPC}.

\section{Results Discussion}\label{sec:results}

\begin{figure*}[h]
  \centering
  \includegraphics[width=0.9\textwidth]{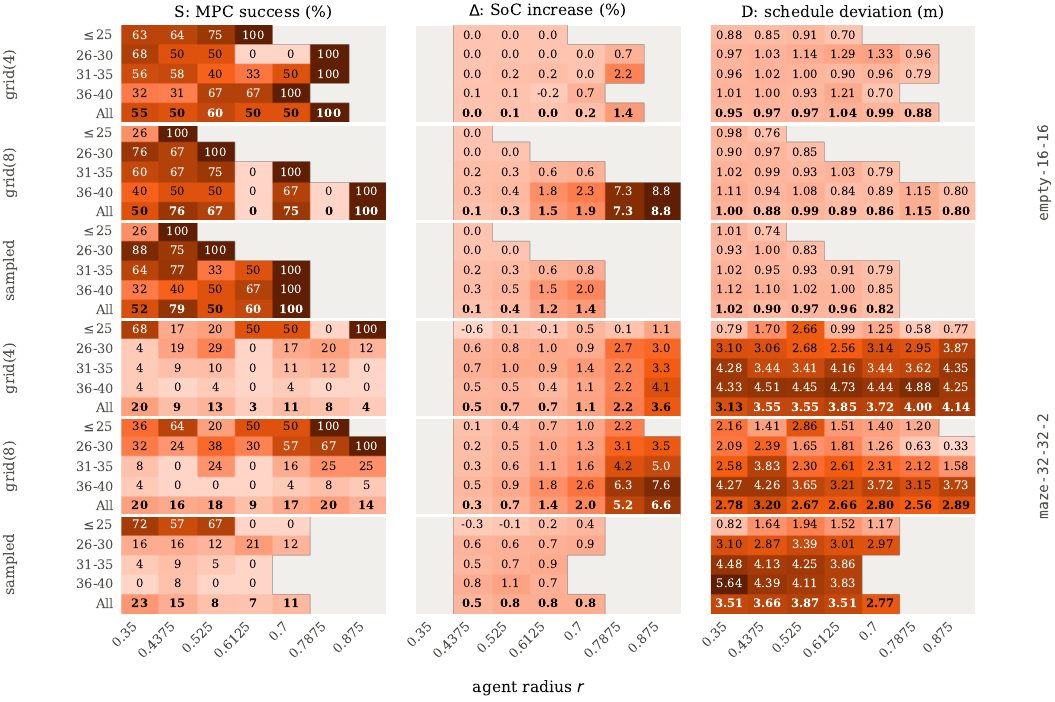}
  \caption{Effect of the agent radius $r$ (no conflict time margin). Panels and conventions as in Figure~\ref{fig:heat_margin}. No scheduler panel is shown: AOC-CBS solved every instance at every radius on both maps. Each larger radius was run only where every smaller one had failed.}
  \Description{}
  \label{fig:heat_radius}
\end{figure*}

The discussion is organized around three questions: how well optimal continuous-time schedules execute as they are (Section~\ref{sec:res:baseline}), what a temporal margin buys and what it costs (Section~\ref{sec:res:temporal}), and whether spatial inflation is an equivalent instrument (Section~\ref{sec:res:spatial}). Throughout, the \emph{scheduler success rate} is the ratio between the instances where AOC-CBS returned a schedule within the 60s budget and the total number of instances; the \emph{MPC success rate} is the ratio between the instances where the MPC drove every agent to its goal without a collision and without exceeding the lateness threshold of 30s, and the total number of instances solved by the scheduler. Figures~\ref{fig:heat_margin}-\ref{fig:heat_radius} report the latter, while the former is discussed in the text.
On the other hand, if an agent arrives more than 30s late to \emph{any} node in its path, it is counted as late; a run is terminated when all agents are either labeled as \emph{finished} after they arrived at their destination or \emph{late}. As long as there there is one agent running that is not late the run continues. Note that a late agent might get delayed, but still manage to complete its path, or get completely stuck in a deadlock. 
In case of lateness, the success rate of the instance is computed as the ratio between \emph{timely} agents (not marked as late) and total agents.
Unless stated otherwise, numbers are aggregated over 21--40 agents and five seeds, and every cost figure $\Delta$ is a per-instance comparison against the same instance's baseline. 
The sweep over different values of time margin and agent radius is performed only for instances that do not fully succeed for a given value of either parameter; for instance, when sweeping over the time margin, if $m=2$ leads to $S < 100 \%$, then the same instance (in terms of number of agents, connectedness, seed, and radius) is run with $m=5$. Otherwise, we assume that if $m=2$ led to $S = 100\%$, so will $m=5$; hence we skip this run. 
AOC-CBS is nondeterministic under its budget; 
in these experiments, repeated identical runs differ 20\% of the time in success rate, and by a median of 0.5\% in cost. 
Where it does not converge, on \maze with sample roadmap for instance, $\Delta$ can turn negative.
% The one exception to this is when AOC-CBS, which does not guarantee a solution under a time budget, was unable to produce a schedule.
% In our experiments, repeated identical runs differ 20\% of the time in success rate and by a median of 0.5\% in cost.
% Cases where no results are shown for higher margins despite not reaching 100\% success rate at the previous margin means that AOC-CBS was unable to produce a schedule within 60s, therefore prohibiting execution. 

\subsection{Optimal Schedules with no Margins}\label{sec:res:baseline}

\begin{figure*}[t]
  \centering
  \includegraphics[width=0.9\textwidth]{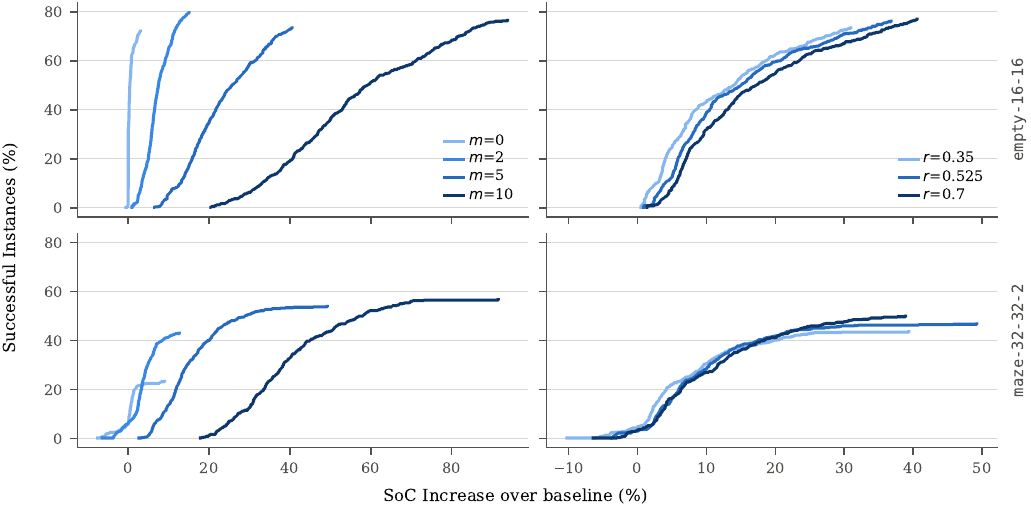}
  \caption{Fraction of instances executed to completion within a given SoC increase relative to the unconstrained baseline. Sweep over time margin on the left, and agent radius on the right. Negative values reflect the anytime budget.}
  \Description{}
  \label{fig:cactus}
\end{figure*}

The baseline configuration with no temporal margin and no spatial inflation confirms that schedule-level optimality is a poor predictor of execution with the MPC tested in this work. On the open map \Empty, between 50\% and 55\% of runs execute to completion depending on the graph abstraction; on \maze the figure falls to 20--23\% (Figure~\ref{fig:heat_margin}, column $m=0$). 
The failures are not dominated by collisions: 3.9\% of baseline runs on \Empty and 7.7\% on \maze end in a collision, while the large majority of failures are lateness failures, in which agents fall far enough behind the schedule that the run is abandoned. This is the expected behavior of a schedule with no slack. 
The scheduler removes separation to minimize the sum of costs, and the controller, which decelerates into turns and tracks its reference with non-zero error, cannot reproduce the timing the schedule assumes. Mean schedule deviation at baseline is 0.95--1.02m on \Empty but 2.78--3.51m on \maze: in a constrained environment the executed trajectories drift significantly, and once that happens the ordering of node traversals that the schedule depends on is disrupted.

\subsection{Temporal Margin Inclusion}\label{sec:res:temporal}

Introducing a conflict time margin changes the picture. On the \maze, raising the margin from $m=2$ to $m=5$ increases success by 28--41 percentage points on the 4-connected grid.
The corresponding increase in sum-of-cost is $\Delta\!\approx\! 17$--21\% (Figure~\ref{fig:heat_double}). Pushing further to $m=10$ continues to help on the \maze, but the sum-of-cost rises to 37--57\% for a much smaller increment in success. 
The frontier therefore has a visible knee: in constrained environments $m=5$ captures most of the available executability for roughly a fifth of the cost, and beyond it the exchange rate deteriorates sharply. Figure~\ref{fig:cactus} makes the point directly: on the \maze, $m=2$ plateaus just above 40\% of instances however generous the cost budget, $m=5$ reaches roughly 50\% at a budget of about 40\%, and $m=10$ reaches the same 50\% only at roughly twice that budget.

The mechanism is visible in the deviation column. On the \maze, mean schedule deviation 
falls sharply with the margin, from 2.78--3.51m at baseline to roughly 0.8--1.4m at $m=10$.
The margin gives the controller the temporal room to avoid accumulating deviation, so the executed trajectory stays close to the scheduled one, and the agents' nodes and edges ordering remain valid.

On \Empty, the benefits are less evident. Success moves from roughly 50\% to the low-to-mid 70s and then flattens, and deviation is already below 1m at baseline. 
Where agents have space to resolve local conflicts by maneuvering, the controller absorbs timing error on its own, and the margin is largely redundant. Which knob is needed is thus a property of the environment, not of the solver.

\subsection{Comparison to Spatial Inflation}\label{sec:res:spatial}

\begin{figure*}[h]
  \centering
  \includegraphics[width=\textwidth]{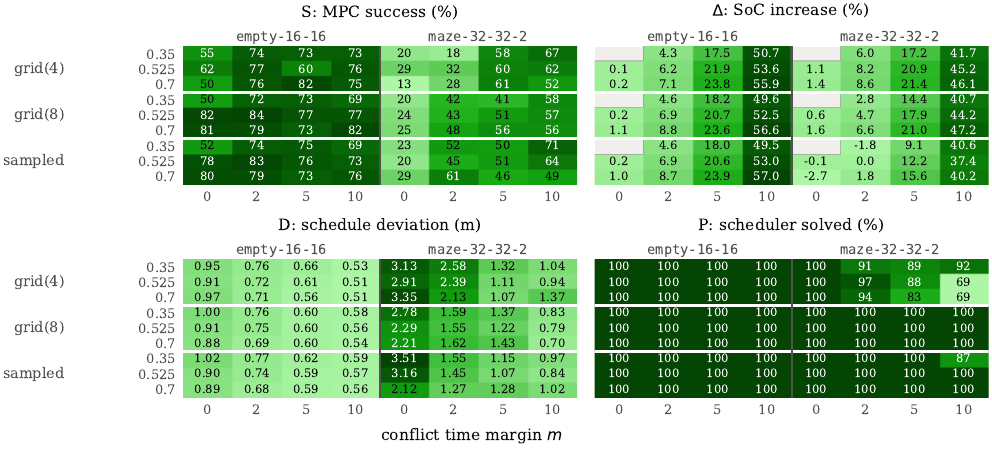}
  \caption{Combined effect of $r$ and $m$, 21--40 agents, seeds 5--9; every cell is fully crossed at 100 instances. Panels and conventions as in Figure~\ref{fig:heat_margin}. The $(r{=}0.35,m{=}0)$ cell is the unconstrained baseline, so its $\Delta$ is blank.}
  \Description{}
  \label{fig:heat_double}
\end{figure*}

The spatial alternative behaves differently. Inflating the footprint from $r=0.35$ to $r=0.7$ costs at most 1.6\% in planned sum of costs (Figure~\ref{fig:heat_double}, $m=0$ columns), against 37--57\% for $m=10$, and it leaves schedule deviation essentially unchanged at a fixed margin. Its effect on execution success, however, is not uniform. Of the 72 paired comparisons between radii at a fixed margin, 6 are significant after correction. Five lie on \Empty and favor the larger radius, four of them at $m=0$: with no temporal margin,
inflation is the only slack available, and it is used. 
The sixth lies on \maze with the sampled roadmap at $m=10$ and favors the \emph{smaller} radius, $r=0.35$ executing successfully on 24 instances where $r=0.7$ fails, against 7 the other way.

% Spatial inflation is not inert, however. Collisions, although rare in absolute terms, fall with radius at a fixed margin on the \maze; with $m=2$, from 4.7\% at $r=0.35$ to 2.7\% at $r=0.7$. 
The temporal margin addresses lateness by preserving the timing the schedule assumes, whereas spatial inflation addresses proximity by ensuring that a given deviation does not translate into contact. 
Because lateness dominates the failure budget in our setting, the temporal margin dominates the overall success rate and the radius increase does not contribute much, actually worsening the result for \maze.

\subsection{The limits of large margins}\label{sec:res:limits}

Margins cannot be increased indefinitely, and the binding constraint is the scheduler rather than the controller. 
On the maze, AOC-CBS solves 92\% of the 100 instances at $m=10$ on the 4-connected grid. Larger margins were run only where a smaller one had failed, so their solve rates are taken over progressively smaller and harder subsets: 71\% of 35 instances at $m=15$, 40\% of 20 at $m=20$ and 23\% of 13 at $m=30$; the \emph{sampled} roadmap degrades similarly. The 8-connected grid is unaffected, remaining at 100\% solved across the whole margin range.
% , which suggests that the richer neighborhood gives the solver enough routing freedom to satisfy a large temporal separation without exhausting the search. Beyond $m\approx 10$, therefore, apparent losses in end-to-end success are increasingly attributable to the planner failing to return a schedule at all, not to execution failures.
The radius compounds this: on the 4-connected maze at $m=10$ the scheduler solves 92\% of instances at $r=0.35$ but only 69\% at $r\geq0.525$. The geometric and computational ceilings therefore reinforce one another.

Two caveats bound these conclusions. First, success increases with the margin but decreases sharply with fleet size: on the \maze, success falls from 33--85\% at 21--24 agents to near zero at 37--40 agents at $m=2$, and the margin shifts the curve without changing its slope. The margin buys executability; it does not buy scalability. 

\section{Conclusions and Future Work}\label{sec:conclusions}

We integrated AOC-CBS, an anytime-optimal continuous-time MAPF solver, into a fleet coordination framework in which every agent tracks its timed path under model predictive control, and we observed that, without a margin, only a minority of instances successfully execute, given a set lateness threshold. Agents instead fall behind their schedules, and the ordering the schedule relies on no longer holds.

Adding a temporal margin repairs this efficiently, but only where it is needed. On the \maze it roughly doubles execution success for a modest increase in scheduled sum-of-costs, and it substantially reduces schedule deviation. Beyond that point, the cost grows much faster than the return, so the frontier has a pronounced knee, and a fleet in a constrained environment should sit at it rather than past it.

Different maps show different results, and the reason is the controller's room to maneuver. On the \Empty, the smallest margin we test already produces a significant change, and larger ones bring little improvement while costing more. Open space gives the MPC somewhere to go: when an agent falls behind, or another crosses its path, the controller can deviate laterally, overtake, or absorb the error and recover the schedule on its own, so slack built into the schedule might be redundant with the controller's own. 
In the maze, that freedom is absent. In a corridor there is no lateral room, an agent that must wait blocks everything behind it, and a local deviation cannot be recovered locally. The slack has to be present in the schedule because the controller cannot compensate at run time. 

Spatial inflation is the same idea applied to a different axis; both instruments buy separation, and either could in principle absorb the controller's tracking error. What separates them is available range, not mechanism. Enlarging the footprint is nearly free in schedule cost, but it leaves schedule deviation unchanged and does not significantly move execution success. The reason is that the spatial axis is bounded by the geometry of the graph: an agent cannot be wider than the corridors it must traverse, and two agents cannot be separated by more than the edges between them allow, so past a certain footprint no schedule exists at all. The temporal axis carries no such bound, since time can always be added. Spatial inflation simply runs out of room before it can supply the slack that lateness demands.

Both instruments therefore have a ceiling, but of different kinds. The temporal one is computational: the bottleneck moves from the controller to the scheduler, which increasingly fails to return any schedule.
% (with of the 8-connected graph in our experiments, whose richer neighborhood leaves more routing freedom). 
The spatial one is geometric, and arrives sooner. 
Whether the two converge on maps with wider corridors, where the geometric ceiling is further away, is an open question. 
Also, it might be more efficient to spend slack more selectively, since only a minority of locations are genuinely contested; letting the solver place margin during search rather than imposing it uniformly should also relieve the timeouts above.
A margin is in any case an open-loop remedy, and combining it with an execution-management layer that reorders agents online would let the schedule carry less slack to begin with, while maintaining executability.
More broadly, these results argue that continuous-time MAPF solvers should be evaluated with a controller in the loop, and that reported optimality should be accompanied by a measure of whether the schedule survives execution.

\newpage

%%%%%%%%%%%%%%%%%%%%%%%%%%%%%%%%%%%%%%%%%%%%%%%%%%%%%%%%%%%%%%%%%%%%%%%%

%%% The acknowledgments section is defined using the "acks" environment
%%% (rather than an unnumbered section). The use of this environment 
%%% ensures the proper identification of the section in the article 
%%% metadata as well as the consistent spelling of the heading.

\begin{acks}
We gratefully acknowledge the Vinnova project EUREKA ITEA4 ArtWork (The smART and connected WORker)
\end{acks}

%%%%%%%%%%%%%%%%%%%%%%%%%%%%%%%%%%%%%%%%%%%%%%%%%%%%%%%%%%%%%%%%%%%%%%%%

%%% The next two lines define, first, the bibliography style to be 
%%% applied, and, second, the bibliography file to be used.
\balance
\bibliographystyle{unsrt}
\bibliography{sample}

%%%%%%%%%%%%%%%%%%%%%%%%%%%%%%%%%%%%%%%%%%%%%%%%%%%%%%%%%%%%%%%%%%%%%%%%

\end{document}